\documentclass[prb,twocolumn,showpacs]{revtex4}

\usepackage{graphpap}
\usepackage[dvips]{graphicx}
\usepackage[dvips]{graphics}
\usepackage{color}

\begin{document}

\title{Time-resolved charge detection in graphene quantum dots}

\author{J. G\"uttinger}
\email{guettinj@phys.ethz.ch}
\author{J. Seif}
\affiliation{Solid State Physics Laboratory, ETH Zurich, 8093 Zurich, Switzerland}
\author{C. Stampfer}
\altaffiliation{Present address: JARA-FIT and II. Institute of Physics, RWTH Aachen, 52074 Aachen, Germany}
\affiliation{Solid State Physics Laboratory, ETH Zurich, 8093 Zurich, Switzerland}
\author{A. Capelli}
\affiliation{Solid State Physics Laboratory, ETH Zurich, 8093 Zurich, Switzerland}
\author{K. Ensslin}
\affiliation{Solid State Physics Laboratory, ETH Zurich, 8093 Zurich, Switzerland}
\author{T. Ihn}
\affiliation{Solid State Physics Laboratory, ETH Zurich, 8093 Zurich, Switzerland}

\date{ \today}
 
\begin{abstract}
We present real-time detection measurements of electron tunneling in a graphene quantum dot. By counting single electron charging events on the dot, the tunneling process in a graphene constriction and the role of localized states are studied in detail. In the regime of low charge detector bias we see only a single time-dependent process in the tunneling rate which can be modeled using a Fermi-broadened energy distribution of the carriers in the lead. We find a non-monotonic gate dependence of the tunneling coupling attributed to the formation of localized states in the constriction. 
Increasing the detector bias above $V_{\mathrm{b}} = 2$~mV results in an increase of the dot-lead transition rate related to back-action of the charge detector current on the dot. 
\end{abstract}


\maketitle

\section{Introduction}
The high sensitivity of a quantum point contact or a single electron transistor to its electrostatic environment is widely used as a powerful tool to detect the electron occupation in semiconductor-based quantum dot structures.\cite{fulton1987, fie93, pet05, ihn09} 
Time resolved charge detection,\cite{lu03,elz04,byl05,fuji06,gus06,fli09} offers the possibility to measure extremely small currents by counting single electron transitions~\cite{byl05} and enables e.g. the extraction of detailed electron tunneling statistics and probing of electron-electron correlations.\cite{fuji06, gus06, fli09} Time-resolution further allows for single shot read out of spin-qubits after spin to charge conversion with potential applications in future quantum information processors.\cite{elz04,hanson07}

With the rise of two-dimensional graphene~\cite{nov04,gei07, neto2009} a fascinating new mesoscopic material for transport experiments has become available with the promise of long spin coherence times.
First nanostructures have been made by etching graphene into narrow constrictions (nanoribbons).\cite{han07,che07} In these devices transport around the charge neutrality point is suppressed. Short constrictions have been successfully used as barriers for graphene quantum dots.\cite{sta08a,pon08,liu09} Several experiments have been performed on graphene quantum dots including the observation of excited states in single~\cite{sch09,mos09} and double quantum dots~\cite{liu10, mol10} and the investigation of orbital~\cite{gue09} and spin states~\cite{gue10} around the electron-hole crossover. An additional nanoribbon placed in close proximity to the dot can be used to detect the number of electrons on the dot~\cite{gue08} (see also Ref.~\onlinecite{hef10}). In contrast to charge detection with a quantum point contact (QPC) tuned below the lowest conductance plateau
, highly charge-sensitive resonances in the detector nanoribbon are used as a sensor, similar to charge detection with a single electron transistor.\cite{fulton1987, lu03, bart10} These resonances arise from localization of charge carriers due to strong potential fluctuations in the disordered nanoribbon.\cite{sol99,tod09,sta09,mol09a,liu09,han10} Here we investigate tunneling through a graphene quantum dot lead in a time-resolved way by counting individual charging events with such an integrated graphene charge detector. The time resolution allows for a deepened analysis of the tunneling properties of a graphene constriction compared to the time-averaged case~\cite{gue08,hef10}.

The paper is organized as follows: we first briefly characterize the charge detector and the quantum dot in Sec.~II. In Sec.~III time-resolved charge detection is presented. It is shown that despite inherent resonances and transmission modulation in graphene barriers, transport can be well understood within a conventional model. For a particular gate-voltage regime an asymmetric double barrier model is successfully used to describe the transmission through a graphene constriction. While increasing the bias, back-action of the detector on the dot is observed. In Sec.~IV the performance of the detector is discussed. A short summary of the results is given in Sec.~V.

\begin{figure}[hbt]\centering
\includegraphics[draft=false,keepaspectratio=true,clip]
                   {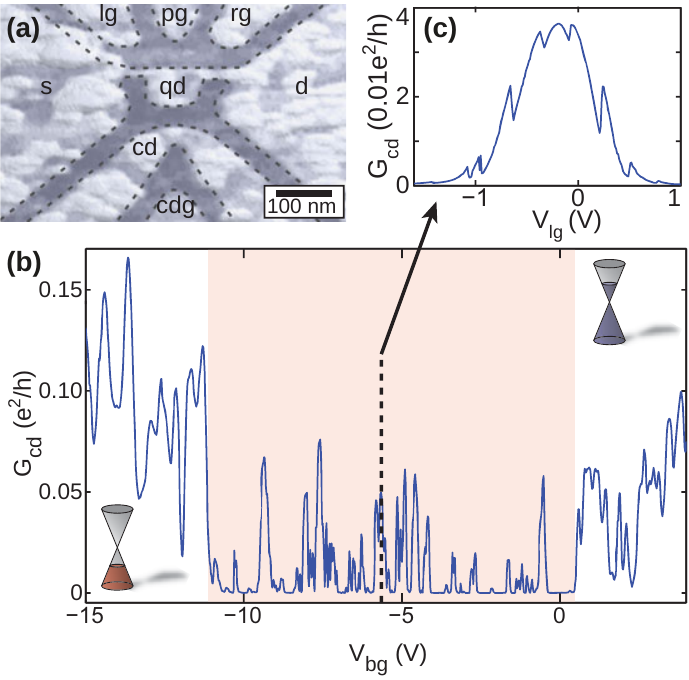}              
\caption[FIG1]{(Color online) (a) SFM micrograph of of the graphene quantum dot with source (s), drain (d) and lateral gates (pg, lg, rg) as well as the charge detector (cd) and its gate (cdg) (b) Measurement of the conductance through the CD as a function of back gate voltage $V_\mathrm{bg}$ at a bias of $V^\mathrm{cd}_\mathrm{b} = 10$\,mV. The cones indicate hole (left) and electron (right) transport, whereas the reddish area marks the transport gap. (c) Single resonance in the charge detector conductance recorded as a function of $V_\mathrm{lg}$ at $V_\mathrm{bg}^\mathrm{eff,cd} = -5.66~$V with $V^\mathrm{cd}_\mathrm{b} = 7$\,mV. The effective $V_\mathrm{bg}$ contains the contribution from the back gate $V_\mathrm{bg} = -2.083~$V and the influence of the charge detector gate $V_\mathrm{cdg} = -11.56~$V (as in the other measurements shown in the paper while all other gates including $V^\mathrm{qd}_\mathrm{b}$ are at 0~V ). The abrupt changes in $G_\mathrm{cd}$ arise from charging the dot with additional electrons.
}
\label{transport}
\end{figure}

\section{Device characterization}
The structure is carved out of a mechanically exfoliated graphene flake by reactive ion etching in an argon-oxygen plasma. Details of the fabrication process can be found in Ref.~\onlinecite{gue09b}. The all-graphene sample [shown in Fig.~1(a)] consists of a 95x70~nm quantum dot (qd) with three lateral graphene gates: left gate (lg), plunger gate (pg) and right gate (rg) [see Fig.~1(a)].
A nanoribbon is used as a charge detector (cd) with one additional lateral charge detector gate (cdg) for tuning the detector to the regime of highest sensitivity. The highly p-doped Si substrate, isolated by 295~nm SiO$_2$, is used as a back gate to tune the Fermi energy. The constrictions connecting the dot to the leads are only 15~nm and 20~nm wide to achieve very low tunneling rates of the order of 1~Hz-100~Hz. The narrower constriction turned out to be completely isolating and no carrier tunneling between dot and source was observed. All measurements were performed in a $^{4}$He cryostat at a base temperature of $T = 1.7$~K.

By changing the back-gate voltage the charge detector can be tuned from hole dominated transport at $V_{\mathrm{bg}}< -11.5~$V to electron dominated transport at $V_{\mathrm{bg}}> 0.5~$V [see Fig.~1(b)]. In between the conductance is pinched off and shows resonances which are the signatures for a "transport gap"~\cite{tod09,sta09,mol09a,liu09} where the conductance is governed by localization and Coulomb blockade induced by disorder.\cite{han10} The center of the gap at $V_\mathrm{bg} \sim 5~$V is offset from $V_{\mathrm{bg}}= 0$ indicating an overall negative doping of the charge detector. 

Fig.~1(c) shows a close-up of a resonance recorded by changing the left gate voltage $V_{\mathrm{lg}}$ while all other gate voltages are fixed in the gate regime indicated by the dashed line (see caption). Due to the high sensitivity of the detector in this gate configuration, this regime is also used in the following measurements. The kinks on the resonance originate from charging the dot electron by electron as $V_{\mathrm{lg}}$ is increased. Whenever an additional electron is loaded to the dot the electrostatic potential in the detector changes abruptly, which results in an abrupt increase (reduction) of the detector conductance if the potential energy is above (below) the resonance condition in the detector. The gate voltage shift $\Delta V_{\mathrm{lg}} \approx 0.13~$V of the curve at each step corresponds to the potential change in the nanoribbon due to the additional dot electron. 

Note that the spatial locations of the charging events are identified by analyzing the influence of the gates on the jumps and resonances in $G_\mathrm{cd}$ (not shown here, see Ref.~\onlinecite{gue08}). The influence is characterized by a lever arm $\alpha_g^{(x)}$ which relates the gate voltage $V_g$ to the induced potential change in device $x$. The result is tabulated in Tab.~1 and shows a similar influence of the right and left gate on the abrupt changes of $G_\mathrm{cd}$ and superior coupling to the plunger gate. Such a behavior is expected if the abrupt changes result from charging the dot. Moreover, the absolute values of the lever arms can be estimated using dot bias spectroscopy by assuming symmetric electrostatic coupling to source and drain, revealing an addition energy of $E_\mathrm{c} = 19$~meV  (not shown). This is in good agreement with the charging energy $E_\mathrm{c}^\mathrm{disk} = e^2/(4\epsilon_0\epsilon d) \approx 20~$meV for an isolated disk of diameter $d=90~$nm, expected (as an upper limit) from the dimensions of the island. 
The spacing of the jumps in gate voltage are then related to an absolute plunger gate lever arm $\alpha_{\mathrm{pg}}^\mathrm{qd} = 0.10$, based on which the other absolute lever arms in Tab.~1 are estimated. 

\begin{table}
	\centering
		\begin{tabular}{|c||c|c|c|c||c|}
		\hline
			relative $\alpha$ & abrupt change & resonance &$\,$ &absolute $\alpha$& $\alpha^\mathrm{qd}$ \\ \hline
			$\alpha_{\mathrm{pg}}/\alpha_{\mathrm{bg}}$& 0.47 & 0.09& &$\alpha_{\mathrm{pg}}$ & $\approx 0.10$ \\ \hline
			$\alpha_{\mathrm{rg}}/\alpha_{\mathrm{bg}}$& 0.28 & 0.07& &$\alpha_{\mathrm{rg}}$ & $\approx  0.06$ \\ \hline
			$\alpha_{\mathrm{lg}}/\alpha_{\mathrm{bg}}$& 0.28 & 0.06& &$\alpha_{\mathrm{lg}}$ & $\approx 0.06$ \\ \hline
			$\alpha_{\mathrm{cdg}}/\alpha_{\mathrm{bg}}$& 0.08 & 0.31& &$\alpha_{\mathrm{cdg}}$ & $\approx 0.02$ \\ \hline
			& & & & $\alpha_{\mathrm{bg}}$ & $\approx 0.21$ \\ \hline
		\end{tabular}
		\caption[Tab1]{Lever arms of the different gates on abrupt quantum dot conductance changes and the resonances in the charge detector. The lever arms are extracted by comparing the influence of each gate relative to the back gate. The absolute plunger gate lever arm on the (dot-) charging events is extracted from dot-bias dependent measurements by assuming symmetric coupling to source and drain. The other absolute dot lever-arms are calculated based on this estimate.}
		\label{tableverarms}
\end{table}

The height of the step $\Delta I_{\mathrm{step}}$ and hence the magnitude of the detector signal depends (i) on the induced change of the detector potential ($e \alpha_{\mathrm{lg}}^\mathrm{cd}\Delta V_{\mathrm{lg}} \sim 1$~meV) given by the electrostatic coupling between the dot and the detector and (ii) on the sensitivity of the detector conductance to potential changes (steepness of the detector current measured as a function of gate voltage). Here, $\alpha_{\mathrm{lg}}^\mathrm{cd}$ is estimated from geometrical considerations to be  $\alpha_{\mathrm{lg}}^\mathrm{cd} \sim \alpha_{\mathrm{cdg}}^\mathrm{qd} \approx 0.02$.

\section{Time resolved charge detection}
\subsection{Time dependent detector current around a step}

\begin{figure}[t]\centering
\includegraphics[draft=false,keepaspectratio=true,clip]%
                   {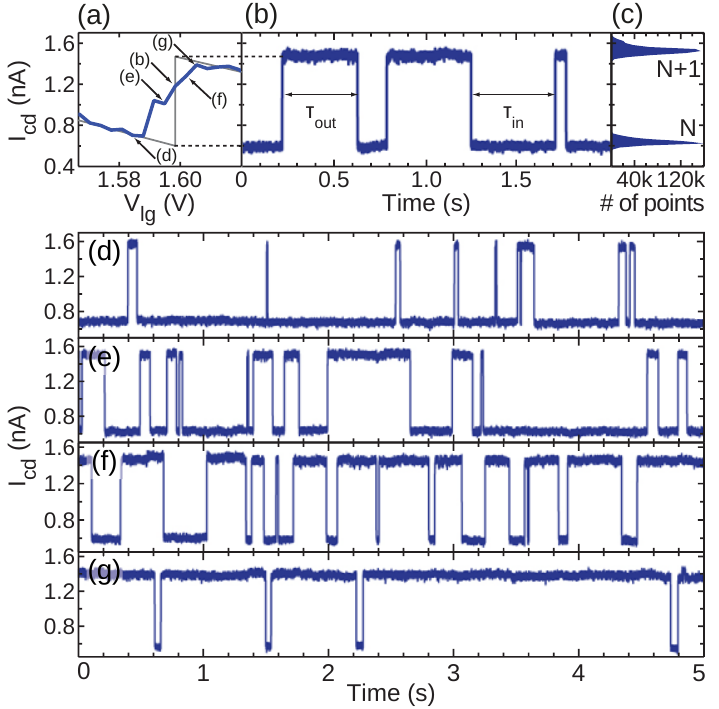}  
\caption[FIG2]{(Color online). (a) Time averaged current through the charge detector as a function of $V_\mathrm{lg}$ while scanning over a similar resonance in the QD as shown in Fig.~1(c). The step indicates a single change from N to N+1 electrons on the dot at $V_\mathrm{b}^\mathrm{cd} = 0.5$~mV. (b) Time resolved current through the charge detector taken at the point labeled with (b) in (a). The time an electron needs to tunnel into or out of the dot is marked with $\tau_{in}$ and $\tau_{out}$ respectively. (c) Histogram of the $I_\mathrm{cd}$ values for the whole time trace of 60~s. (d-g) Time traces taken at different $V_\mathrm{lg}$ marked in (a), show a gradual change of the dot electron number from N (lower level) to N+1 electrons (upper level). 
} 
\label{experiment}
\end{figure}

In the following we analyze the detector conductance around such a step in real time. Fig.~2(a) shows a zoom of the time averaged current across a step measured with a typical integration time of 0.2~s. Here a lower bias of $V_\mathrm{b}^\mathrm{cd} = 0.5~$mV is applied. 
We don't observe a single step (see gray line) but the signal is noisy and the transition is smeared out. These are indications that the timescale of the transition is comparable to the measurement integration time. Indeed, by measuring a time trace at position (b) [see Fig.~2(b)] we observe a two-level random signal switching in intervals of around 0.5~s. 
The two-level signal shows a large signal to noise ratio (SNR) of $\Delta I_{\mathrm{step}}/\left\langle I_{\mathrm{noise}}\right\rangle \approx 30$ at a measurement system bandwidth of $\approx 400$~Hz. The noise $\left\langle I_{\mathrm{noise}}\right\rangle$ is defined as the variance $\sqrt{\mathrm{Var}(I)}$ of the detector current on each of the two current levels. The two levels indicate whether there are $N$ electrons (lower) or $N+1$ electrons (higher level) on the dot and hence allow real time detection of single charge carriers tunneling on and off the dot. The corresponding dwell times $\tau_{\mathrm{in}}$ and $\tau_{\mathrm{out}}$ are indicated in the figure. The histogram in Fig.~2(c) shows the distribution of the current for a 60~s time trace with a clear separation of the two states. The slight asymmetry in the distribution of the upper level is attributed to an additional weakly coupled charge fluctuator present in the $N+1$ state. 

By tuning the dot potential away from the resonance condition the occupation probability of the lower $N$ [upper $N+1$] level is reduced as seen in Fig.~2(e,d) [f,g]. In Fig.~2(g) for example the chemical potential for the $N+1$ transition lies below the chemical potential in the lead. Therefore the dwell time for the empty state $\tau_{\mathrm{in}}$ is much smaller than $\tau_{\mathrm{out}}$.

\subsection{Quantitative analysis of time traces}

The following analysis follows closely recent work on similar systems realized in semiconductors.\cite{lu03, schle04, gus09review}
\begin{figure}[t]\centering
\includegraphics[draft=false,keepaspectratio=true,clip]
                   {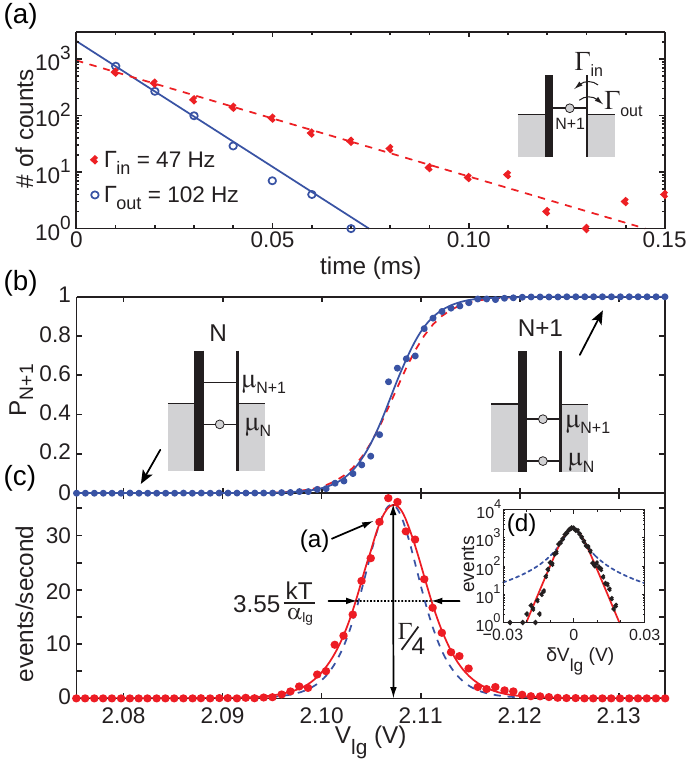}                   
\caption[FIG3]{(Color online). (a) Distribution of the dwell times recorded at a position where the dot chemical potential is slightly above the drain chemical potential [see schematic and indicated by the arrow in (c)]. The distribution shows an exponential behavior described in Eq.~\ref{eq1} with $\Gamma_\mathrm{in} = 47$~Hz and $\Gamma_\mathrm{out} = 102$~Hz. (b) Occupation probability for the N+1 electrons dot state $P(N+1)$ as a function of $V_\mathrm{lg}$. The solid line is a fit to $1-f(E)$ with f the Fermi function. Assuming $T = 1.7$~K a $V_\mathrm{lg}$ lever arm of $\alpha_\mathrm{lg} = 0.077 e$ is obtained. For comparison the dashed line shows the result obtained in (c). Here the events per second are plotted for varying $V_\mathrm{lg}$. Comparing these event rates with Eq.~(\ref{eq2}) the tunneling coupling and the lever arm are given as $\Gamma = 143$~Hz and $\alpha_\mathrm{lg} = 0.067 e$ (solid line). The difference of the lever arms extracted from (b) and (c) (dashed line) is attributed to the limited statistics. (d) The data from (c) is fitted in addition with a tunnel-coupling broadened Lorentzian line shape $ \propto \Gamma/\left(\alpha_\mathrm{lg}(\delta V_\mathrm{lg})^2+\Gamma^2\right)$ (dashed line) and plotted in logarithmic scale, confirming the thermal broadening of the resonance. This measurement has been conducted at a dot resonance close to the one analyzed in Fig.~2 with faster tunneling rates [see Fig.~4(a)].
} 
\label{fig3}
\end{figure}
A quantitative analysis of time traces reveals information about the number of dot-levels participating in transport, the tunneling coupling, the carrier temperature and distribution in the leads and the individual tunneling rates. For this analysis the barrier is slightly opened to obtain larger count rates and therefore improved statistics. In Fig.~3(a) the distribution of the dwell times is plotted for a situation where the chemical potential of the dot state $\mu_{N+1}$ is slightly above the chemical potential of the drain $\mu_{\mathrm{d}}$ (see schematic). The dwell times are exponentially distributed indicating that only a single dot level is involved in the transport process. Hence the probability density $p_{\mathrm{in/out}}(t)$ which is the number of counts with dwell time $\tau_{\mathrm{in/out}} = t$ normalized by the total number of counts is given by
\begin{equation}
p_{\mathrm{in/out}}(t)dt = \frac{1}{\left\langle \tau_{\mathrm{in/out}}\right\rangle} \exp{\left(-\frac{t}{\left\langle \tau_{\mathrm{in/out}}\right\rangle}\right)}dt.
\label{eq1}
\end{equation}
From a fit of this equation to the data in Fig.~3(a) tunneling rates $\Gamma_{\mathrm{in}} = \frac{1}{\left\langle \tau_{\mathrm{in}}\right\rangle} = 47$~Hz and $\Gamma_{\mathrm{out}} = \frac{1}{\left\langle \tau_{\mathrm{out}}\right\rangle} = 102$~Hz are extracted. 

As shown in Fig.~2(d-g) these rates change by tuning the potential of the dot. Fig.~3(b) shows the occupation probability of the N+1 state changing from 0 to 1 while lowering the chemical potential of the dot for increasing $V_{\mathrm{lg}}$ (see schematics). If we assume gate voltage-independent tunneling coupling to the lead the occupation probability is determined by the distribution of the charge carriers in the lead. 

We assume a Fermi distribution in the lead $f = [1+\exp{(\Delta\mu/kT_{\mathrm{e}})}]^{-1}$ with $\Delta\mu$ the difference between the chemical potential of the dot and the lead. 
By tuning the left gate voltage $\Delta\mu$ is changed according to $\Delta\mu = e\alpha_{\mathrm{lg}}(V_\mathrm{lg}-V_\mathrm{res})$ where $\alpha_{\mathrm{lg}}$ is the lever arm of the left gate on the dot (the small lever arm on the lead is neglected). The occupation probability is then given as $P_{N+1}(\Delta\mu) = 1-f(\Delta\mu)$ and we can extract $\alpha_{\mathrm{lg}} \approx 0.077\pm 0.01$ under the assumption that the electron temperature $T_{\mathrm{e}}$ is equal to the bath temperature of 1.7~K. The uncertainty arises from the limited statistics around the crossover point and is reduced by analyzing the number of tunneling events (see below). The lever arm is comparable to $\alpha_\mathrm{lg} \approx 0.06$ estimated from the dot bias dependence in Sec.~II.

Combining single-level transport and constant tunnel coupling it is possible to extract the tunnel coupling $\Gamma$ by counting the number of tunneling events [see Fig.~3(c)]. 
The event rate $r_{\mathrm{e}}$ for tunneling-in is given by
\begin{equation}
r_{\mathrm{e}} = \frac{1}{\left\langle \tau_{\mathrm{in}}\right\rangle + \left\langle \tau_{\mathrm{out}}\right\rangle} = \Gamma \cdot f(1-f).
\label{eq2}
\end{equation}
The best fit to the data yields $\Gamma = 143$~Hz and $\alpha_{\mathrm{lg}} \approx 0.067\pm0.005$. The fit obtained in (b) is plotted for comparison as a dashed line in (c) and vice versa in (b). The discrepancy between the two fits is explained by the larger influence of the data points around the resonance in (b) compared to (c). 

In Fig.~3(d), a Lorentzian broadened line shape (dashed) and the thermally broadened fit (solid) of the number of events are plotted on a logarithmic scale for comparison. As expected for low tunneling coupling the thermally broadened line shape describes the data much more accurately.


\subsection{Changing the tunneling barrier}

\begin{figure}[t]\centering
\includegraphics[draft=false,keepaspectratio=true,clip]
                  {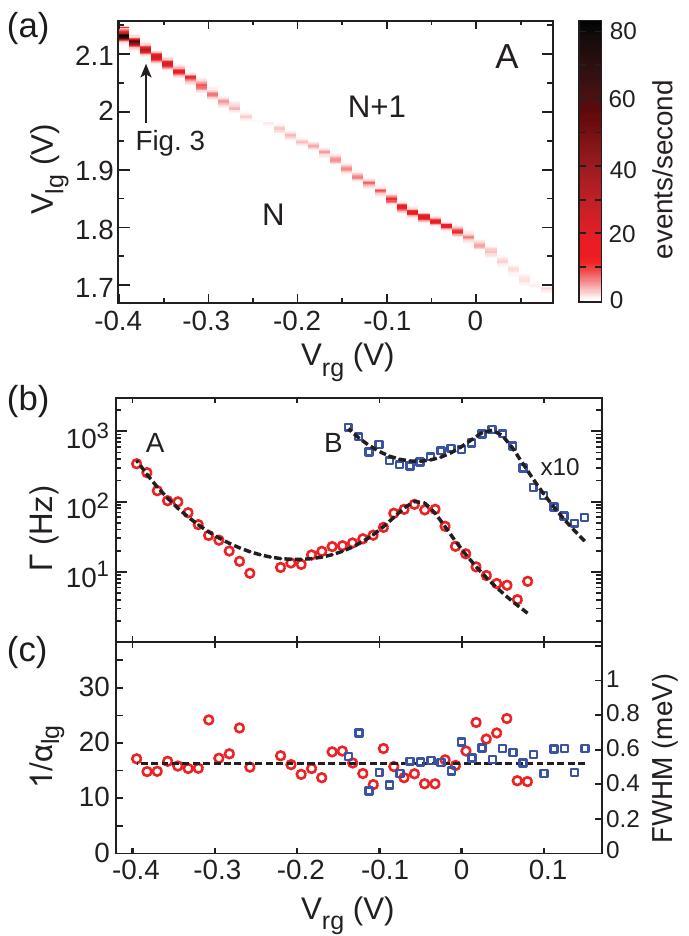}                   
\caption[FIG4]{(Color online). (a) Number of events per second measured as a function of $V_\mathrm{lg}$ and $V_\mathrm{rg}$ while crossing a dot resonance. A single trace taken at $V_\mathrm{rg} = -0.37~$V has been analyzed in Fig.~3. (b) Tunneling coupling as a function of $V_\mathrm{rg}$ (and $\sim -V_\mathrm{lg}$) plotted in logarithmic scale. The red circles are obtained from the data shown in (a) (regime A), whereas the blue squares are deduced from a similar measurement in a slightly shifted gate regime B around $V_\mathrm{rg} = 0$~V and $V_\mathrm{lg} = 1.75$~V after a small charge rearrangement (10x magnified for clarity). The non-monotonic behavior can be explained by modeling the constriction as asymmetric double barrier with lorentzian shaped resonances from the weak barrier ($\hbar\Gamma_\mathrm{A} = 6.2~$meV and $\hbar\Gamma_\mathrm{B} = 6.4~$meV) and an exponential suppression due to the strong barrier (see text for details). (c) Peak width plotted as inverse lever arm $1/\alpha_\mathrm{rg}$ and FWHM (right scale) as a function of $V_\mathrm{rg}$ in both regimes with an average $\alpha_\mathrm{rg} = 0.062$. Here $V_\mathrm{b}^\mathrm{cd} = 0.5$~mV and the counting time is 60~s per point. 
} 
\label{fig4}
\end{figure}

By changing the corresponding side gate we can tune the barrier and the tunnel coupling.\cite{sta08a,mol09b} Fig.~4(a) shows the number of events per second as a function of $V_{\mathrm{lg}}$ and $V_{\mathrm{rg}}$. The measurement analyzed in Fig.~3 is a cut at $V_{\mathrm{rg}} = -0.37~$V. The diagonal line corresponds to the resonance condition where an additional carrier is loaded onto the dot (N $\rightarrow$ N+1). The potential of the dot is affected equally by both gates as expected from the geometry [Fig.~1(a)]. In addition a change in the number of counts is observed especially at lower $V_{\mathrm{rg}}$. This change in the barrier transmission $\Gamma$ can be seen more easily in Fig.~4(b) where the corresponding tunneling coupling $\Gamma$ [obtained by fitting to Eq.~(\ref{eq2})] is plotted as a function of $V_{\mathrm{rg}}$ (A, red circles). The same behaviour is also observed in a second measurement (B, blue squares) taken in the same gate regime after a small charge rearrangment and offset with a factor of 10 for clarity. While the tunneling coupling varies strongly by changing $V_{\mathrm{rg}}$ the full-width-half-maximum (FWHM), which is inversely proportional to the lever arm is approximately constant with an average $\alpha_\mathrm{lg} = 0.062$ at $T = 1.7~$K [see Fig.~4(c)]. This averaged lever arm is in good agreement with the $\alpha_\mathrm{lg} \approx 0.06$ obtained by varying the dot bias. 

Unlike GaAs based quantum dots where tunneling rates tend to increase monotonically with gating due to depletion of the electron gas~\cite{lean07}, the non-monotonic exponential changes of the tunneling rate in our graphene nanostructure is an indication for the presence of resonances in the constrictions.\cite{sta08a, mol09b, sch10} It is important to note that we see no charging of constriction resonances in this measurement. However, it is possible to tune the device into a regime where the typical hexagon pattern of a double dot is measured while changing $V_{\mathrm{rg}}$ versus $V_{\mathrm{lg}}$~\cite{wiel02} and signatures of a second localized state in the barrier can be observed. If charging of well localized parasitic resonances is slow enough even additional small steps in the counting signal are observable in those regimes.

In the gate regime investigated in this paper, the influence of localized states on the dot energy is negligible but still the barrier transmission is modulated. Such a behavior might occur if the additional localized state is strongly coupled to the lead but only weakly to the dot. This situation can be modeled with a one-dimensional asymmetric double barrier with tunneling coupling $\Gamma_R \gg \Gamma_L$. For noninteracting electrons in the case of $kT \ll h\Gamma \ll \Delta$ (with $\Delta$ the level spacing) the total transmission is given as~\cite{sto85,buet86,buet88}
\begin{equation}
T_\mathrm{tot} = \sum_p\Gamma_\mathrm{L}\frac{\Gamma_\mathrm{R}}{(\Gamma_\mathrm{R}/2)^2 + (E_F-E_p)^2}
\end{equation}
if we assume $\Gamma_\mathrm{L,R}$ to be independent of the resonance $p$. In this limit the Lorentzian shape is caused by the weak barrier with strong coupling while the overall amplitude is determined by the weak tunneling coupling of the strong barrier. For the strong barrier we assume an exponential dependence of $\Gamma_\mathrm{L}$ on the gate voltage while the gate dependence of the weak barrier is neglected. The dashed lines in Fig.~4(b) are the corresponding fits with two resonances ($p = 1,2$). The extracted tunneling coupling of the weak barrier is similarly strong in both measurements with $\Gamma_\mathrm{R,A} = 6.2~$meV and $\Gamma_\mathrm{R,B} = 6.4~$meV assuming a typical lever arm of $\alpha_\mathrm{rg,loc} = 0.1$ on the localized state. For the left barrier the WKB result with a linearized exponential $\Gamma_\mathrm{L} = \Gamma_0 \exp{[-\kappa(\delta U- \delta E)]}$ is used.~\cite{lean07} The details of the barrier are described by $\Gamma_0$ and $\kappa$ while $\delta E$ ($\delta U$) describes small perturbations of the dot energy (barrier potential). By keeping the dot energy constant and assuming a linear gate dependence of the barrier potential we get $\Gamma_\mathrm{L} = \Gamma_0 \exp{[-\beta V_\mathrm{rg}]}$ where $\beta = \kappa\alpha_\mathrm{rg}^\mathrm{barrier}$. For the two measurements we obtain $\Gamma_{0,A} = 0.1~$Hz, $\beta_A = 6~$V$^{-1}$ and  $\Gamma_\mathrm{0,B} = 0.2~$Hz, $\beta_B = 11~$V$^{-1}$.
The different parameters in the two regimes are attributed to a change of the left barrier potential between the two measurements.

\begin{figure}[t]\centering
\includegraphics[draft=false,keepaspectratio=true,clip]
                  {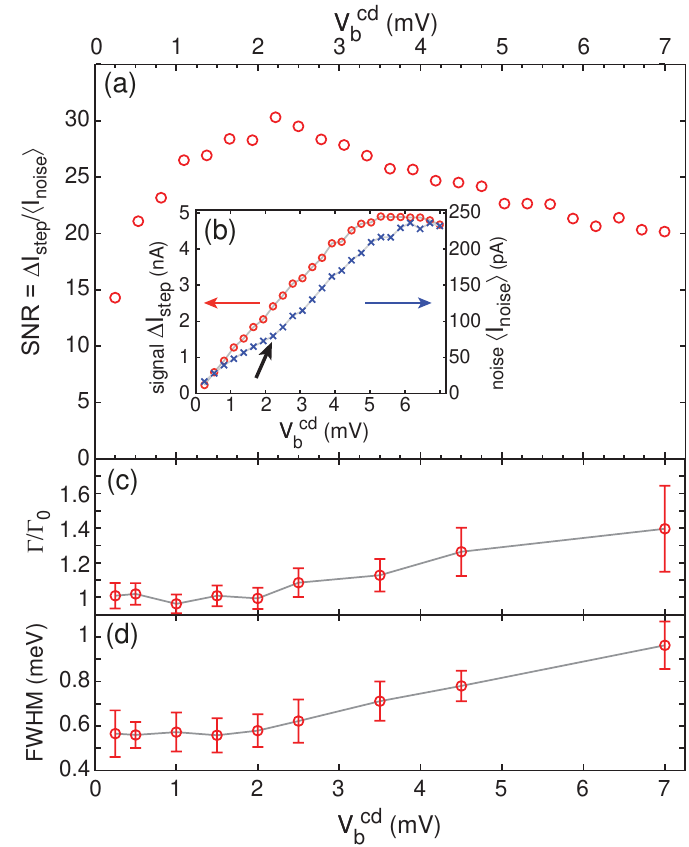}                   
\caption[FIG5]{
(Color online). (a) Signal to noise ratio for increasing charge detector bias $V_\mathrm{b}^\mathrm{cd}$. Every data point is obtained from the average step height (signal) and the average noise on the two levels from 60 time traces ($V_\mathrm{lg} = 1.735-1.79$~V) each 25 seconds long at $V_\mathrm{rg} = 0$. (b) Signal and noise versus $V_\mathrm{b}^\mathrm{cd}$ plotted independently. The bold arrow marks the onset of the stronger increase in noise giving rise for the saturation of the signal to noise ratio in (a). (c,d) Dependence of the dot events on the detector bias. Here the measurement B shown in Fig.~4(b,c) for $V_\mathrm{b}^\mathrm{cd} = 0.5~$mV is repeated for different values of $V_\mathrm{b}^\mathrm{cd}$ with $-150$~mV$ < V_\mathrm{rg} < 25~$mV. In (c) the relative change of the tunneling rate $\Gamma$ is plotted. In order to compensate for variations in the tunneling rate by changing $V_\mathrm{rg}$, $\Gamma_0$ is defined as the average tunneling rate for $V_\mathrm{b}^\mathrm{cd} < 2~$mV for each $V_\mathrm{rg}$.
In (d) the FWHM of the peaks [as shown in Fig.~3(c)] averaged over the different $V_\mathrm{rg}$ is plotted for increasing $V_\mathrm{b}^\mathrm{cd}$. 
} 
\label{fig5}
\end{figure}

\subsection{Detector bias dependence and back-action}
In the measurements shown so far the bias in the charge detector was kept low ($V_{\mathrm{b}}^{\mathrm{cd}} = 500~\mu$V) in order to prevent back-action of the detector on the dot.\cite{buk98,spr00} On the other hand a higher charge detector bias leads to an increase of the signal (-step). In order to maximize the performance, the SNR is investigated as a function of charge-detector bias in Fig.~5(a,b). The SNR is maximized for  $V_{\mathrm{b}}^{\mathrm{cd}}= 2~$mV and gets smaller for  $V_{\mathrm{b}}^{\mathrm{cd}}> 2~$mV due to a stronger increase of the noise [bold arrow in Fig.~5(b)], as can be seen in Fig.~5(b), where the signal and the noise are separately plotted. This higher noise is correlated with an increase in tunneling events and a broadening of the dot event peak [see Fig.~5(c,d)]. The data is obtained by recording a left-gate right-gate map of the charging events [such as Fig.~4(a)] for different $V_{\mathrm{b}}^{\mathrm{cd}}$. The extracted FWHM of the peak and tunneling rates are averaged over 15 right gate values ($V_\mathrm{rg} = [-150 -25~$mV]) in regime B (see squares in Fig.~4(b,c) where the tunneling coupling and the FWHM are shown as a function of $V_\mathrm{rg}$ at $V_{\mathrm{b}}^{\mathrm{cd}} = 0.5~$mV). In order to account for variations of the tunneling rate with $V_\mathrm{rg}$, an average tunneling rate $\Gamma_0(V_\mathrm{rg})$  for $V_{\mathrm{b}}^{\mathrm{cd}}< 2~$mV is defined for each $V_\mathrm{rg}$ value. In Fig.~5(c) the average rate $\Gamma/\Gamma_0$ increases up to 40\% from $V_{\mathrm{b}}^{\mathrm{cd}}\leq 2~$mV to $V_{\mathrm{b}}^{\mathrm{cd}}= 7~$mV. Note also the increase of the standard deviation of the average value, reflected in the errorbar. In Fig.~5(d) the FWHM is calculated from the peak width using the leverarm $\alpha_\mathrm{lg} = 0.06$ obtained from Fig.~4(c). Similar to the tunneling rate the peak width depends  approximately linearly on the detector bias above $V_{\mathrm{b}}^{\mathrm{cd}}= 2~$mV.

Due to the correlation with the noise in the charge detector, we attribute the back-action from the detector on the dot to arise mainly from shot noise generated in the detector constriction.\cite{buk98,gus07} Photon emission and absorption is rather easy in graphene due to the linear, zero-bandgap electronic dispersion. Heating due to acoustic phonons~\cite{khr06,gas09} is less plausible because the phonons have to couple via the SiO$_2$ substrate over a different material~\endnote{The maximum detector current of $I_{\mathrm{cd}} = 22~$nA  at $V_{\mathrm{b}}^{\mathrm{cd}}= 7~$mV ($I_{\mathrm{cd}} = 4~$nA at $V_{\mathrm{b}}^{\mathrm{cd}}= 2~$mV) is rather small compared to the change in FWHM corresponding to a change in $T_\mathrm{el}$ from $1.7$~K to $3.1$~K. In Ref.~\onlinecite{gas10} a GaAs-QPC heated the reservoirs of a double dot by 1~K with a current of $I_\mathrm{qpc} \approx 160$~nA.}.
In addition, graphene has a very high thermal conductivity ($\approx 5000~$W/m/K)~\cite{bala08} 
compared to SiO$_2$ (1.3 W/K/m, both measured at $T = 300~$K) and therefore the generated heat is expected to thermalize in the leads of the graphene constriction rather than to heat the dot lead via the oxide. However, in order to clarify this mechanism further experiments including double quantum dots with frequency resolved absorption measurements~\cite{agu00} or more involved studies of the detector-current dependence at constant detector bias are desirable.

\section{Detector performance} 
In the presented device a current step of up to 2~nA with a relevant noise spectrum below $2~\mathrm{pA}/\sqrt{\mathrm{Hz}}$ is measured at a detector bandwidth of 4~kHz ($V_{\mathrm{b}}^{\mathrm{cd}}= 1.7~$mV) (see Fig.~6). This corresponds to a charge sensitivity of the detector better than $10^{-3}e/\sqrt{\mathrm{Hz}}$ comparable to what has been reported in Ref.~\onlinecite{van04apl} for a GaAs QPC detector.

\begin{figure}[t]\centering
\includegraphics[draft=false,keepaspectratio=true,clip]
                  {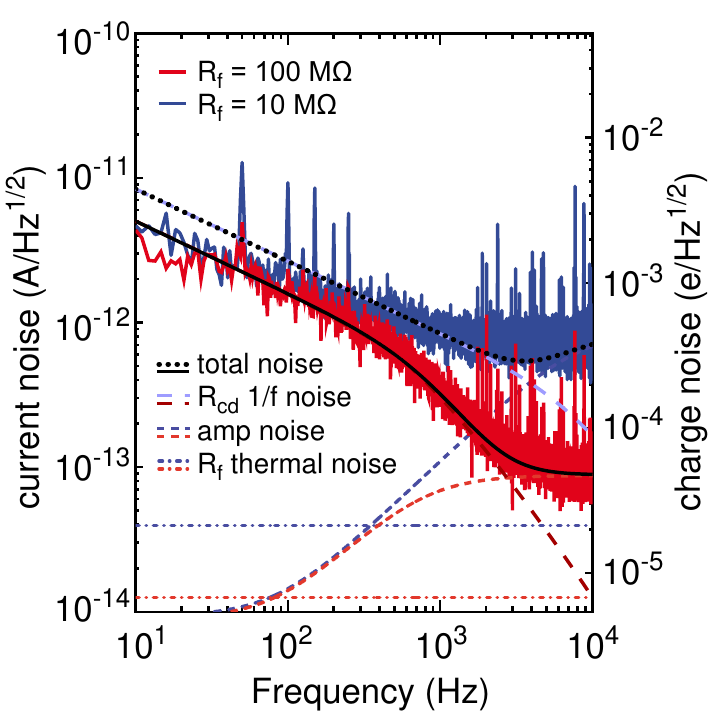}                   
\caption[FIG6]{
(Color online). (a) Current- and charge-noise versus frequency for different feedback resistors $R_\mathrm{f} = 10~$M$\Omega$ (blue, bandwidth 4~kHz) and 100~M$\Omega$ (red, bandwidth 400~Hz) measured at a bias $V_{\mathrm{b}}^{\mathrm{cd}}= 1.7~$mV with optimal signal-to-noise ratio ($R_\mathrm{cd} = 570~$k$\Omega$). The noise is obtained by averaging the noise spectra of 10 time-traces, each 0.5~s long (containing no dot charging events). The total modeled noise (black) is mainly determined by 1/f-noise of the sample (long dashes) and the amplifier noise (short dashes). The thermal noise of the feedback resistor is lower in magnitude (dash-dotted). On the right scale the charge noise is plotted. It is obtained by dividing the noise by the step height of the signal $I_\mathrm{step} = 1.9~$nA induced by charging the dot with an additional electron. The value below $10^{-3}e/\sqrt{\mathrm{Hz}}$ above $f=1$~kHz is comparable with the resolution shown in Ref.~\onlinecite{van04apl}. 
}       
\label{fig6}
\end{figure}

Although in principle a bandwidth of up to $30$~kHz can be achieved with a conventional room temperature amplifier setup, as has been shown with GaAs QPC detectors,\cite{elz04,gus06,gus09review} the bandwidth is limited in the presented measurements to below 1~kHz to ensure sufficient SNR for counting. 

The first limitation is posed by the noise of the system. Usually the noise spectrum of a charge detector in such a setup is dominated by intrinsic 1/f-noise from the sample at frequencies $f < 1~$kHz and by amplifier noise at higher frequencies. In our sample the 1/f-noise of the device is roughly three times larger compared with GaAs QPC detectors.~\cite{van04apl,tmu10} 
The negative influence on the SNR is limited by reducing the bandwidth with a larger feedback resistance. Concerning the amplifier noise at higher frequencies, the comparably high resistance of our detector ($R_\mathrm{cd} = 500$~k$\Omega$ compared to $R_\mathrm{cd}^{\mathrm{qpc}} = 35$~k$\Omega$ in QPC-based detectors~\cite{van04apl,gus09review}) leads to a weaker signal amplification and hence the amplifier noise gets proportionally more important. Here an increase of the feedback resistance is beneficial as well, as the amplification is restored. 

A second limitation for the SNR is related to the general difficulty using SET-based charge detectors to maintain the working point with optimal sensitivity. In principle it is possible to compensate the influence of any gate on the charge detector with the charge-detector gate. However, the change of $V_\mathrm{cdg}$ induced additional charge fluctuations in the detector and was therefore kept at a constant value during measurements. 

A more general issue for time resolved charge detection in graphene is the limited tunability of the dot barriers. Although the current through etched graphene quantum dots can be tuned over several orders of magnitude the tunability of the tunneling barrier is limited by the width of the constriction. 

Despite the mentioned issues with the current implementation, we could show high sensitivity with signal changes of more than 50\% at moderate cryogenic temperatures of $T = 1.7~$K. This is because graphene offers the possibility for very strong electrostatic coupling between charge detector and quantum dot, because the spacing between them can in principle be made even smaller than the 30~nm used in this device. The coupling could be even further increased by making use of the monoatomic thickness of graphene and vertically stacking dot and detector.\cite{gus08apl} 
In addition, the bandwidth can be improved using a low temperature amplifier~\cite{lee89,vink07} and/or a radio frequency setup,~\cite{schoel98, lu03, muel07, rei07, cas07} where a bandwidth of 1~MHz has been shown with a charge sensitivity of $\leq 2\cdot 10^{-4}e/\sqrt{\mathrm{Hz}}$ in Ref.~\onlinecite{lu03}.

\section{Summary} 
We demonstrated time-resolved charge carrier detection in a graphene quantum dot with high charge sensitivity of the detector due to its close proximity to the quantum dot. For the measurements recorded at low detector bias, the tunneling rate can be modeled conventionally by a single time-dependent process with a temperature broadened energy distribution of carriers in the lead. Gating of the tunneling barrier reveals a non-monotonic gate dependence of the tunneling coupling by counting individual charging events. This behavior is attributed to resonant tunneling through localized states in the barrier strongly coupled to the lead with $\hbar\Gamma \sim 6~$meV. For detector bias $V_{\mathrm{b}}^{\mathrm{cd}}> 2~$mV we see a symmetric broadening of the energy distribution of the tunneling events accompanied by an increase in the detector noise. This back-action is attributed to shot noise in the charge detector, as graphene offers a high thermal conductivity in contrast to SiO$_2$ together with the ease of photon emission and absorption in an a priori zero-bandgap material. 

The authors wish to thank B.~K\"ung, T.~M\"uller, S.~Gustavsson, P.~Studerus, C.~Barengo and S.~Ludwig for help and discussions.
Support by the ETH FIRST Lab, the Swiss National Science Foundation and NCCR nanoscience are gratefully acknowledged.

\end{document}